\documentclass[aps,prl,reprint,amsmath,amssymb,superscriptaddress]{revtex4-2}
\usepackage[svgnames,psnames]{xcolor}
\usepackage{graphicx,hyperref}
\hypersetup{
colorlinks=true,
citecolor=NavyBlue,
linkcolor=DarkRed
}
\usepackage{bm,ulem}

\begin{document}

\title{Thin Films of Topological Nodal Line Semimetals as a Candidate for Efficient Thermoelectric Converters}

\author{Masashi {\sc Hosoi}}
\thanks{hosoi@hosi.phys.s.u-tokyo.ac.jp}
\affiliation{Department of Physics, University of Tokyo, 7-3-1 Hongo, Bunkyo, Tokyo 113-0033, Japan}
\author{Ikuma {\sc Tateishi}}
\affiliation{Department of Physics, University of Tokyo, 7-3-1 Hongo, Bunkyo, Tokyo 113-0033, Japan}
\affiliation{RIKEN Center for Emergent Matter Science, Wako, Saitama 351-0198, Japan}
\author{Hiroyasu {\sc Matsuura}}
\affiliation{Department of Physics, University of Tokyo, 7-3-1 Hongo, Bunkyo, Tokyo 113-0033, Japan}
\author{Masao {\sc Ogata}}
\affiliation{Department of Physics, University of Tokyo, 7-3-1 Hongo, Bunkyo, Tokyo 113-0033, Japan}

\date{\today}

\begin{abstract}
Thermoelectric materials intrigue much interest due to their wide range of application such as power generators and refrigerators. The efficiency of thermoelectric materials is quantified by the figure of merit, and a figure greater than unity is desired. To achieve this, a large Seebeck coefficient and low phonon thermal conductivity are required. We show that this can be achieved with a thin film of topological nodal line semimetals. We also discusses the correlation effect and spin current induced by a temperature gradient. The obtained results provide insight for the improvement of thermoelectric materials.
\end{abstract}

\maketitle

\textit{Introduction.}
The investigation of materials capable of efficient thermoelectric conversion is central to the development of sustainable energy solutions \cite{chenRecentDevelopmentsThermoelectric2003,heAdvancesThermoelectricMaterials2017}. In particular, global environmental problems mean there is urgent need for materials that can efficiently generate electricity from waste heat. Consequently, there has been considerable effort in the search for materials with a high figure of merit ($ZT$) (see Eq. \eqref{ZT}), which is used to quantify the efficiency of thermoelectric materials, and numerous theoretical studies have discussed its physical limits \cite{slackPropertiesSemiconductingIrSb1994,salesFilledSkutteruditeAntimonides1996,mandrusFilledSkutteruditeAntimonides1997,chungCsBi4Te6HighPerformanceThermoelectric2000,hsuCubicAgPbmSbTe2Bulk2004,tanNonequilibriumProcessingLeads2016,hicksEffectQuantumwellStructures1993,hicksThermoelectricFigureMerit1993,bayerlTheoreticalLimitsThermoelectric2015}. Such materials
include bismuth telluride alloys, which are widely used in thermoelectric devices operating below 500 K and have $ZT$ of $\sim$1 at room temperature \cite{goldsmidBismuthTellurideIts2014}, and  lead telluride alloys, which have been reported with a high $ZT$ of $\sim$2.2 at 915 K \cite{peiHighThermoelectricFigure2011,peiConvergenceElectronicBands2011,biswasHighperformanceBulkThermoelectrics2012}. However, the use of tellurium is problematic because of its toxicity and rarity. Recently, many candidate materials with large $ZT$ that do not contain tellurium have been found \cite{hinterleitnerThermoelectricPerformanceMetastable2019,byeonDiscoveryColossalSeebeck2019}, but the discovery of more promising materials is still highly anticipated.

One strategy to improve thermoelectric performance involves lowering the dimension. Thus, thin films of thermoelectric materials offer an ideal arena for investigation of such low-dimensional effects \cite{venkatasubramanianThinfilmThermoelectricDevices2001,harmanQuantumDotSuperlattice2002,harmanNanostructuredThermoelectricMaterials2005}. To date, two mechanisms have been proposed to explain the enhancement of $ZT$ in thin films. The first mechanism involves the quantum size effect, which can increase the density of states (DOS) around the Fermi energy \cite{dresselhausNewDirectionsLowDimensional2007,maoSizeEffectThermoelectric2016}. When the thickness of the film is comparable to the effective de Broglie wavelength of electrons, their motion is restricted to the two-dimensional plane, yielding a remarkable change in the DOS.
Mahan and Sofo suggested that a sharp and large DOS peak would improve the Seebeck coefficient \cite{mahanBestThermoelectric1996}, and many experiments have supported this prediction \cite{harmanHighThermoelectricFigures1996,hicksExperimentalStudyEffect1996a,harmanPbTeTeSuperlattice1999,ohtaGiantThermoelectricSeebeck2007,wangEnhancedThermopowerPbSe2008}. For example, in high-quality Pb$_{1-x}$Eu$_x$Te/PbTe there are multiple quantum wells and $ZT>$ 1.2, which is several times larger than the bulk value \cite{harmanHighThermoelectricFigures1996}. A recent study showed that $ZT>$ 2.4 is expected for a high-density two-dimensional electron gas confined in a unit-cell-thick layer of SrTiO$_3$ \cite{ohtaGiantThermoelectricSeebeck2007}. The second mechanism involves the phonon-blocking/electron-transmitting nature of superlattices. To maximize $ZT$, the phonon thermal conductivity should be largely suppressed by an acoustic mismatch between the superlattice components. This effect was confirmed experimentally for p-type Bi$_2$Te$_3$/Sb$_2$Te$_3$ superlattices where $ZT$ was $\sim$2.4 at 300 K \cite{venkatasubramanianThinfilmThermoelectricDevices2001}.

In the present paper, we show that these two factors, a sharp DOS peak near the Fermi energy and reduced lattice contributions to the thermal current, are both satisfied by topological semimetals.

Beginning with Weyl and Dirac semimetals, topological semimetals have been extended to include those with line nodes, namely topological nodal line semimetals (TNLSs) \cite{burkovTopologicalNodalSemimetals2011,phillipsTunableLineNode2014,fangTopologicalNodalLine2015,wengTopologicalNodelineSemimetal2015,yuTopologicalNodeLineSemimetal2015,neupaneObservationTopologicalNodal2016,huEvidenceTopologicalNodalLine2016,bianTopologicalNodallineFermions2016,hirayamaTopologicalDiracNodal2017}.
Following the discovery of ZrSiS and PbTaSe$_2$, the first material realizations of TNLSs, many other materials such as the Ca$_2$As and Ag$_2$S families have been examined \cite{neupaneObservationTopologicalNodal2016,bianTopologicalNodallineFermions2016,tateishiNodalLinesMapping2020,huangTopologicalNodallineSemimetal2017}.
A notable property of TNLSs is the possession of drumhead surface states. The nature of their dispersion is less dispersive than those of bulk bands, and their two-dimensionality yields a sharp and large DOS peak near the Fermi energy. Furthermore, these surface states are expected to be robust against disorder because they are topologically protected. Thus, the phonon thermal conductivity could be reduced by disorder in the bulk without changing the electron contributions. In this paper, we study the thermoelectric transport properties of TNLS thin films to find an extremely large $ZT$ greater than ten. In addition, we discuss the possibility of surface magnetism and spin currents \cite{liuCorrelationEffectsQuantum2017,ogataTheorySpinSeebeck2017}.

Topological semimetals such as Dirac and Weyl semimetals have been considered unsuitable for the high thermoelectric performance since they have zero DOS even for the surface states. However, the situation is completely different in TNLSs where the drumhead surface states give large DOS at the Fermi enrgy. As for topological insulators, although a relatively large $ZT$ has been observed in Heusler compounds, there have been no studies of their thermoelectric performance focusing on the specialty of the surface state.

%%%%%%%%%%%%%%%%%%%%%%%%%%%%%%%%%%%%%%%%%%%%
\begin{figure}[h]
\rotatebox{0}{\includegraphics[width=\linewidth]{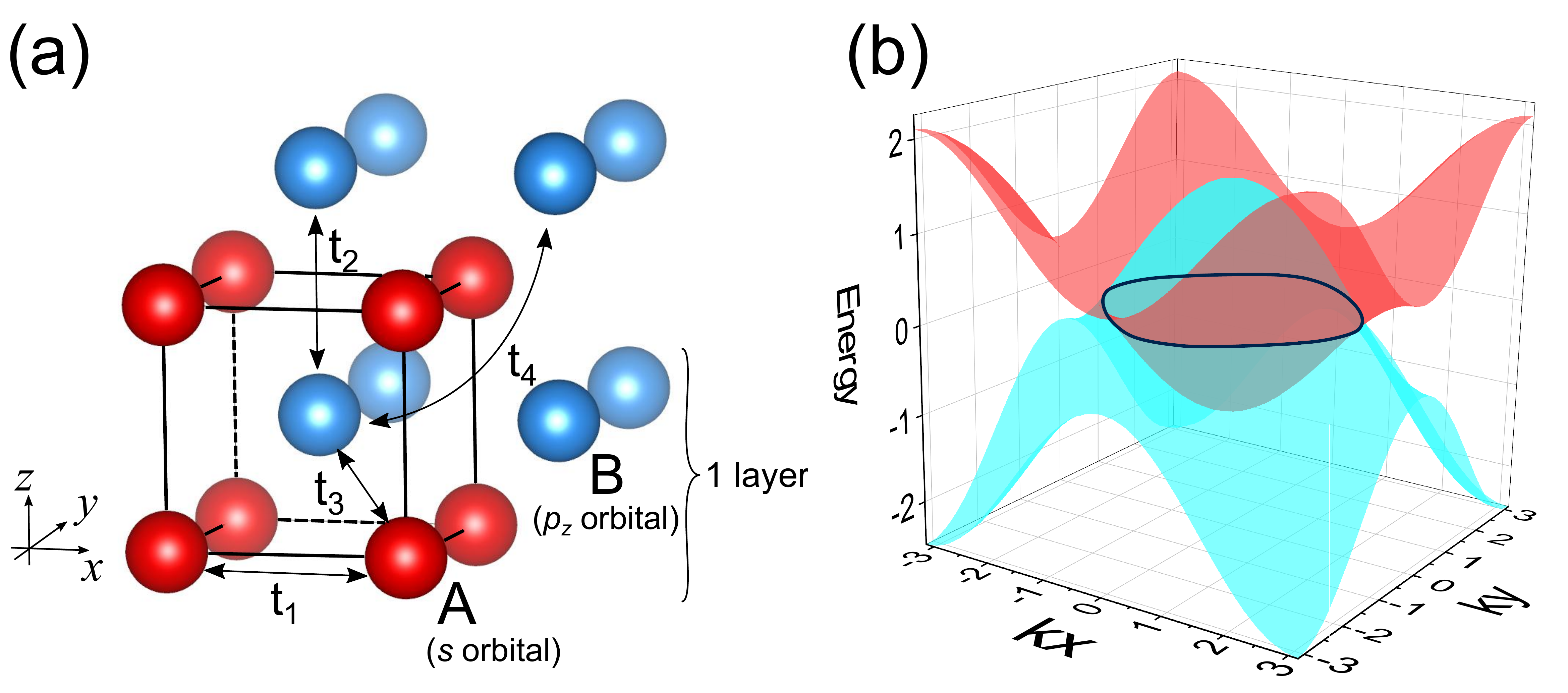}}
\caption{(a) Schematic diagram of the tight-binding model on a primitive tetragonal lattice for TNLSs. Red (blue) spheres represent the A (B) sublattices with $s$ ($p_z$) orbitals. Four different transfer integrals are denoted by $t_1$–$t_4$. (b) Band dispersions of the bulk Hamiltonian at $k_z=0$ with $E_0^A=-E_0^B=-1.5\mbox{, }t_1=0.5\mbox{, }t_2=-0.5\mbox{, }t_3=0.3,$ and $t_4=-0.2$. The nodal line is depicted by the solid black line.}
\label{Fig1}
\end{figure}
%%%%%%%%%%%%%%%%%%%%%%%%%%%%%%%%%%%%%%%%%%%%%

\textit{Model.}
We construct a simple tight-binding model on a primitive tetragonal lattice as an example of a TNLS. Similar results with large Seebeck coefficients would be obtained for any TNLS. As shown in Fig. \ref{Fig1}(a), we assume that each primitive cell hosts two sublattices, A and B, which possess $s$ and $p_z$ orbitals, respectively. In addition, four types of transfer integrals $t_1$--$t_4$ are considered. The Hamiltonian is given by
\begin{equation}
  \label{model}
  \begin{aligned}
    H&=\sum_{i,\sigma,\alpha=A,B}E_0^\alpha c_{i\sigma}^{\alpha\dagger} c_{i\sigma}^\alpha+\sum_{\langle i,j\rangle,\sigma}(t_1c_{i\sigma}^{A\dagger} c_{j\sigma}^A+t_2c_{i\sigma}^{B\dagger} c_{j\sigma}^B+\mathrm{h.c.})\\
    &+t_3\sum_{\langle i,j\rangle,\sigma}(\eta_{ij}c_{i\sigma}^{A\dagger} c_{j\sigma}^B+\mathrm{h.c.})+t_4\sum_{\langle\langle i,j\rangle\rangle,\sigma}(c_{i\sigma}^{B\dagger} c_{j\sigma}^B+\mathrm{h.c.}).
  \end{aligned}
\end{equation}
Here, $c_{i\sigma}^{\alpha\dagger}(c_{i\sigma}^\alpha)$ is the creation (annihilation) operator for electrons in the $\alpha$ sublattice with spin $\sigma$ in the $i$-th unit cell, and $E_0^\alpha$ is the one-body potential of each sublattice. The sign factor $\eta_{ij}=1 (-1)$ if sites $i$ and $j$ belong to the same (different) layer due to the symmetry requirement. Note that the transfer integral $t_4$ is not required to describe TNLSs. This term is introduced to make each band dispersive because, when $t_4=0$, one of the bands becomes exactly flat over the entire Brillouin zone. Hereafter, we set $E_0^A=-E_0^B=-1.5\mbox{, }t_1=0.5\mbox{, }t_2=-0.5\mbox{, }t_3=0.3,$ and $t_4=-0.2$. Moreover, the energy scales are given in eV.

In this tight-binding model, the nodal line resides on the $k_x$--$k_y$ plane (Fig. \ref{Fig1}(b)). Therefore, the drumhead surface state is expected to appear on the (001) surface. To examine its effect on thermoelectric transport, we consider thin films with two different truncations. In type I truncation, the top surface is terminated by a layer of sublattice B, and in type I\hspace{-1 pt}I it is terminated by a layer of sublattice A. In both cases, the bottom surface is terminated by a layer of sublattice A. Hereafter, we focus on type I truncation, and type I\hspace{-1 pt}I truncation, which does not have a surface state, is discussed in the Supplemental Materials. Figure \ref{Fig2} shows the band structure of a 10-layer film along the high-symmetry points of the Brillouin zone and its DOS. This shows that a drumhead surface state is realized, and that its flatness leads to a sharp peak in the DOS near the Fermi energy.
%%%%%%%%%%%%%%%%%%%%%%%%%%%%%%%%%%%%%%%%%%%%%%%%%%%%%%%%%%
\begin{figure}
\rotatebox{0}{\includegraphics[width=\linewidth]{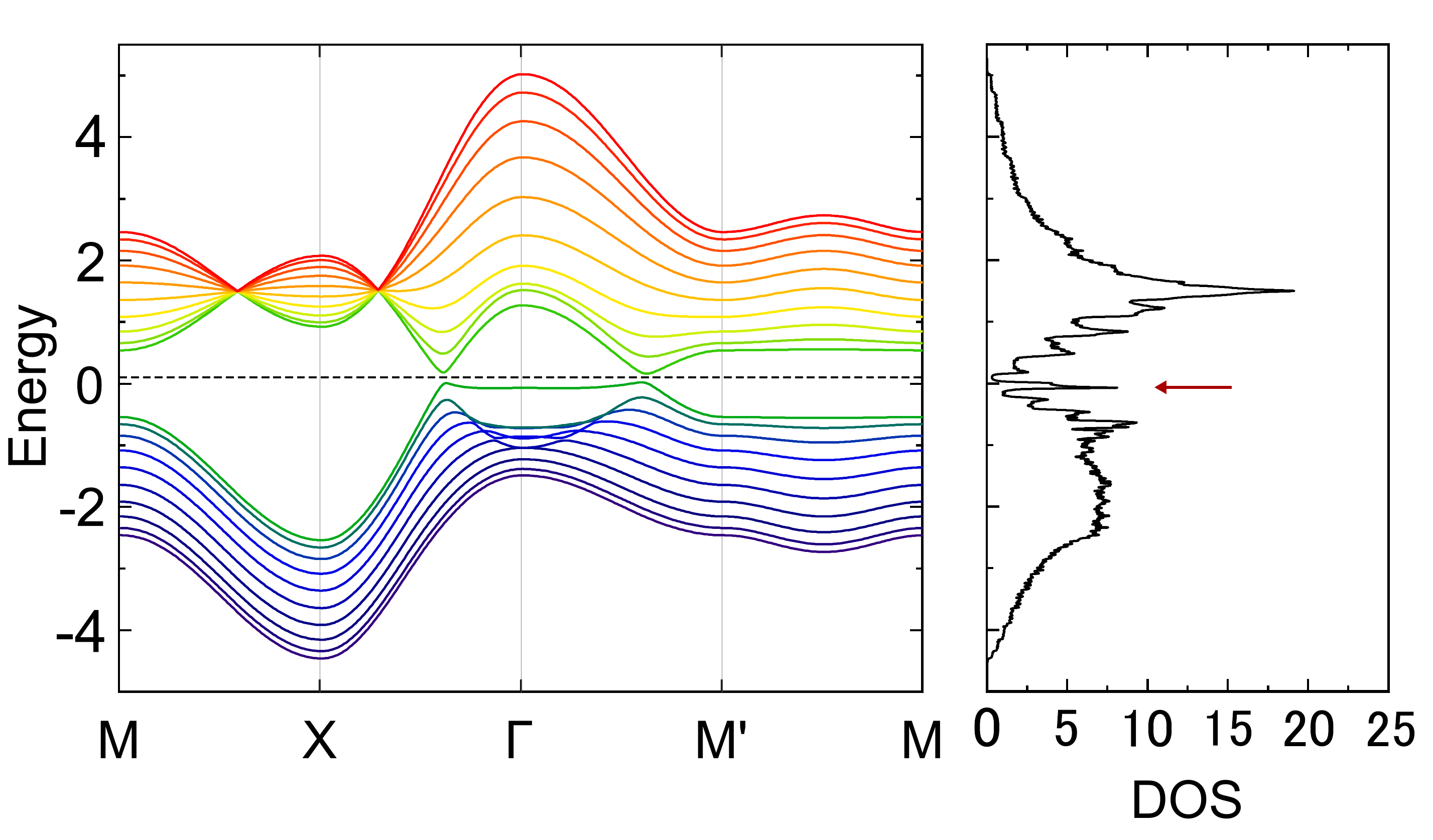}}
\caption{Band structures of a 10-layer film of TNLS with type I truncation along the high symmetry points of the Brillouin zone. The coordinates ($k_x, k_y$) are defined as $\Gamma=(0, 0)$, $\mathrm{M}=(\pi, 0)$, $\mathrm{M}'=(0, \pi)$, and $\mathrm{X}=(\pi, \pi)$. The dashed lines indicate the positions of the Fermi energy. The right panel shows the DOS, and the red arrow indicates a sharp peak supported by a drumhead flat surface state.}
\label{Fig2}
\end{figure}
%%%%%%%%%%%%%%%%%%%%%%%%%%%%%%%%%%%%%%%%%%%%%%%%%%%%%%%%%%%

\textit{Linear response theory.}
The linear responses of the electrical current ${\bm J}_e$ and thermal current ${\bm J}_Q$ to the electric field ${\bm E}$ and temperature gradient $\nabla T$ are given by ${\bm J}_e=L_{11} {\bm E}+L_{12}\left(-{\nabla T}/{T}\right)$ and ${\bm J}_Q=L_{21} {\bm E}+L_{22}\left(-{\nabla T}/{T}\right)$.
Here, $L_{ij}$ are the transport coefficients and $L_{11}$ is the electric conductivity $\sigma$. The Seebeck coefficient $S$ and thermal conductivity $\kappa_e$ owing to electrons are given by
\begin{equation}
  S=\frac{L_{12}}{TL_{11}}=\frac{L_{12}}{T\sigma},
\end{equation}
and
\begin{equation}
\quad\kappa_e=\frac{1}{T}\left(L_{22}-L_{12}L_{11}^{-1}L_{21}\right),
\end{equation}
respectively. The efficiency of thermoelectric energy converters is evaluated by a dimensionless figure of merit defined as
\begin{equation}
  \label{ZT}
ZT=\frac{S^2\sigma}{\kappa}T,
\end{equation}
where $\kappa$ is the total thermal conductivity of the contributions from electrons $\kappa_e$ and phonons $\kappa_{\mathrm{ph}}$. In the following calculations, we only consider $\kappa_e$, so our $ZT$ is the maximum (or most optimistic) value. However, in TNLSs, the surface state is expected to be robust against disorder except on the surface of the material. Therefore, if we introduce disorder in the bulk, we can reduce $\kappa_{\mathrm{ph}}$, while other values remains stable because the surface state is unchanged. This can justify neglecting $\kappa_{\mathrm{ph}}$ in $ZT$.

In the simplest approximation, the constant relaxation time approximation, the electric conductivity along the $x$ direction is given by
\begin{equation}
  \label{sigma}
    L_{11}=\frac{4\tau e^2}{\hbar^2V}\sum_{n,{\bm k}}\frac{\partial \epsilon^n_{{\bm k}}}{\partial k_x}\frac{\partial \epsilon^n_{{\bm k}}}{\partial k_x}(-f'(\epsilon^n_{{\bm k}})),
\end{equation}
where $n$ is the band index, $\tau$ is the relaxation time, $e (<0)$ is the electron charge, and $f(\epsilon)=1/(e^{(\epsilon-\mu)/k_BT}+1)$ is the Fermi distribution function with chemical potential $\mu$. In the thin film case, the summation over ${\bm k}$ is in a two-dimensional space.
To calculate the thermoelectric transport coefficients, we use the following relations derived from the Boltzmann equation \cite{Mahan,SommBethe,MottJones,Wilson}.
\begin{equation}
  \label{SB}
  \begin{aligned}
    L_{12}&=\frac{4\tau e}{\hbar^2V}\sum_{n,{\bm k}}\frac{\partial \epsilon^n_{{\bm k}}}{\partial k_x}\frac{\partial \epsilon^n_{{\bm k}}}{\partial k_x}(-f'(\epsilon^n_{{\bm k}}))(\epsilon^n_{{\bm k}}-\mu),\\
    L_{22}&=\frac{4\tau}{\hbar^2V}\sum_{n,{\bm k}}\frac{\partial \epsilon^n_{{\bm k}}}{\partial k_x}\frac{\partial \epsilon^n_{{\bm k}}}{\partial k_x}(-f'(\epsilon^n_{{\bm k}}))(\epsilon^n_{{\bm k}}-\mu)^2.
  \end{aligned}
\end{equation}
These relations are microscopically justified when we do not consider the heat current of phonons originating from phonon drag \cite{jonsonElectronphononContributionThermopower1990,kontaniGeneralFormulaThermoelectric2003,ogataRangeValiditySommerfeld2019}, and when the relaxation rate $\Gamma=\hbar/2\tau$ is very small compared to $\epsilon^n_{\bm k}$.

We directly calculate the energy dispersions for TNLS films by the exact diagonalization method, and then estimate thermoelectric transport coefficients from Eqs. \eqref{sigma} and \eqref{SB}.
%%%%%%%%%%%%%%%%%%%%%%%%%%%%%%%%%%%%%%%%%%%%%%%%%%%%%%%%%

\begin{figure}[h]
\rotatebox{0}{\includegraphics[width=\linewidth]{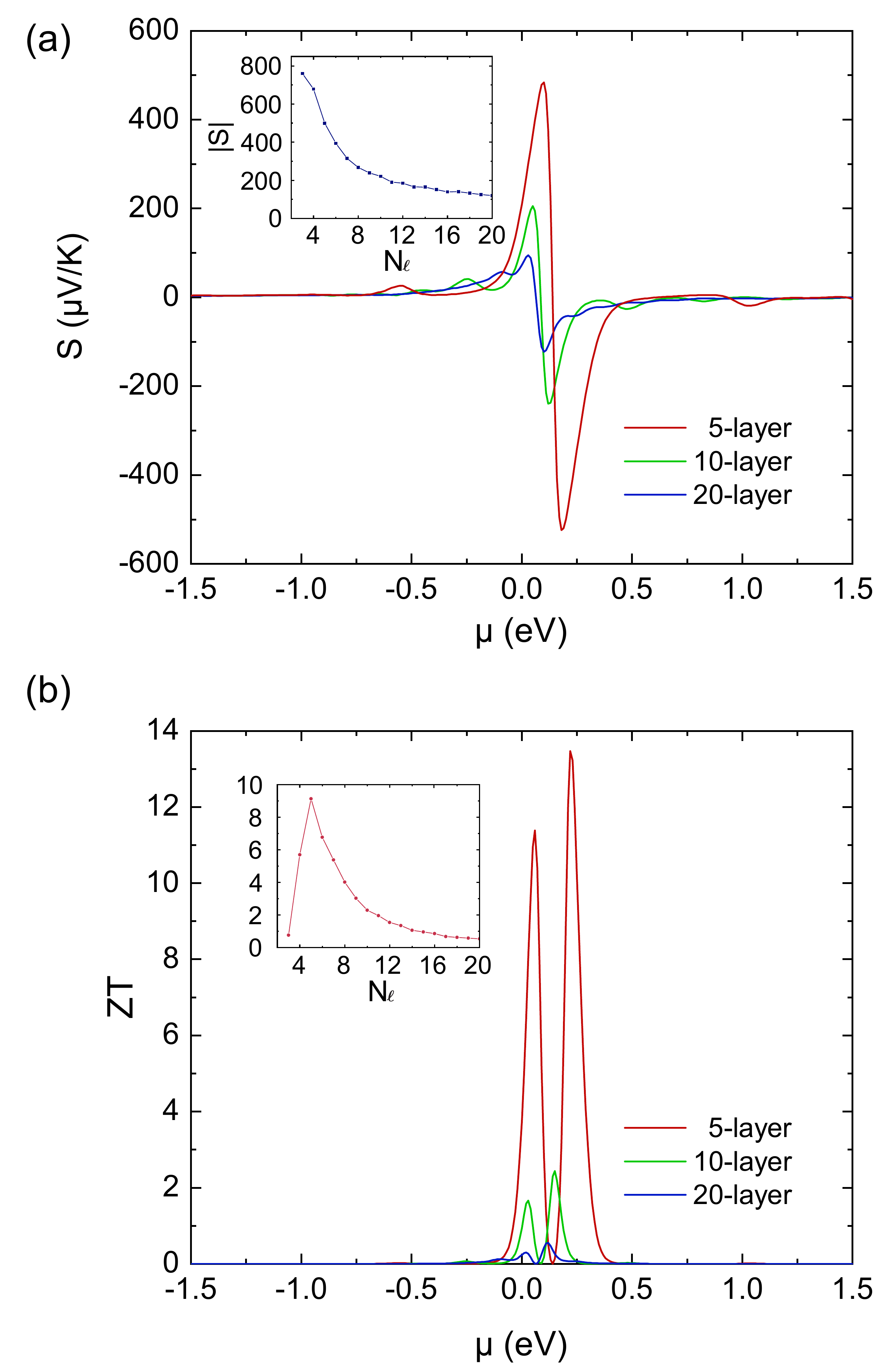}}
\caption{Chemical potential dependence of the thermal coefficients in thin films of TNLS at approximately room temperature ($k_BT=2.5\times10^{-2}$). (a) Seebeck coefficient. (b) Figure of merit. Red, green, and blue solid lines represent 5-, 10-, and 20-layer films with type I truncation, respectively. (Inset) Layer-number ($N_\ell$) dependence of $S$ and $ZT$, where the chemical potential is set to the Fermi energy.}
\label{Seebeck}
\end{figure}

\textit{Thermoelectric transport properties of TNLSs.}
Figure \ref{Seebeck}(a) shows the chemical potential dependence of the Seebeck coefficient in TNLS films with various thicknesses at around room temperature, $k_BT=2.5\times10^{-2}$. The red, green, and blue lines represent the results for 5-, 10-, and 20-layer films with type I truncation, respectively. A large Seebeck coefficient can be observed near the Fermi energy, and it increases as the films get thinner (see the inset, where $N_\ell$ is the number of layers).
Figure \ref{Seebeck}(b) shows the corresponding {$ZT$} with $\kappa_{\mathrm{ph}}$ neglected. The values of $ZT$ are very large, exceeding 13 for the type I 5-layer film, which is five times larger than that of Bi$_2$Te$_3$/Sb$_2$Te$_3$ superlattices. Note that the peak position is almost identical to the Fermi energy of the half-filled system.

When $\mu=0$, the sharp DOS peak due to the drumhead surface state is less than the chemical potential, so $S$ is positive (hole-like). As $\mu$ increases, $S$ become negative (electron-like), probably because of the DOS peak above the chemical potential. As shown in the Supplemental Materials, in the type I\hspace{-1 pt}I truncation case where there are no drumhead surface states, $S$ shows similar behavior as a function of $\mu$, but the absolute values are small. This means that the existence of the drumhead surface state strongly enhances the absolute values of $S$ and $ZT$.

The dependence on the number of layers $N_\ell$ is shown in the inset, where the chemical potential is set to the Fermi energy. The Seebeck coefficient $S$ (inset of Fig. \ref{Seebeck}(a)) monotonically decreases, and $ZT$ (inset of Fig. \ref{Seebeck}(b)) peaks at $N_\ell=5$, then decreases as the film thickness increases. The behavior of $N_\ell\geq5$ is approximated by $S\sim N_\ell^{-1}$ and $ZT\sim N_{\ell}^{-2}$. These results can be roughly explained by the band structures for each thickness. In thin films, the DOS peak near the Fermi energy is conspicuous compared to that in thick films because the DOS originating from bulk bands is not very large. The sharp DOS peak contribute to the large Seebeck effect and the extremely large figure of merit. However, when the film is too thin (i.e., $N_\ell<5$), the finite-size gap becomes large and the surface state is indistinguishable from the bulk bands. Consequently, the conductivity becomes approximately zero and therefore $ZT$ decreases.

%%%%%%%%%%%%%%%%%%%%%%%%%%%%%%%%%%%%%%%%%%%%%%%%%%%%%%%%%%

\textit{Surface magnetism and spin current.}
Next, we discuss the effect of the electron correlation. The DOS at the surface of type I films is large, so Coulomb interaction is highly screened. Thus, on-site Hubbard type interaction, $H_U=\sum_{i,\alpha=A,B}U_\alpha n_{i\uparrow}^\alpha n_{i\downarrow}^\alpha$ describes the effects of the correlation well. Here,  $n_{i\sigma}^\alpha=c_{i\sigma}^{\alpha\dagger}c_{i\sigma}^{\alpha}$ is the number operator. In this work, we evaluate the effects of Hubbard interaction using an unrestricted Hartree-Fock approximation. Quantum fluctuation, not considered here, is expected to reduce magnetic moments and not change the results qualitatively.

\begin{figure}[h]
\rotatebox{0}{\includegraphics[width=\linewidth]{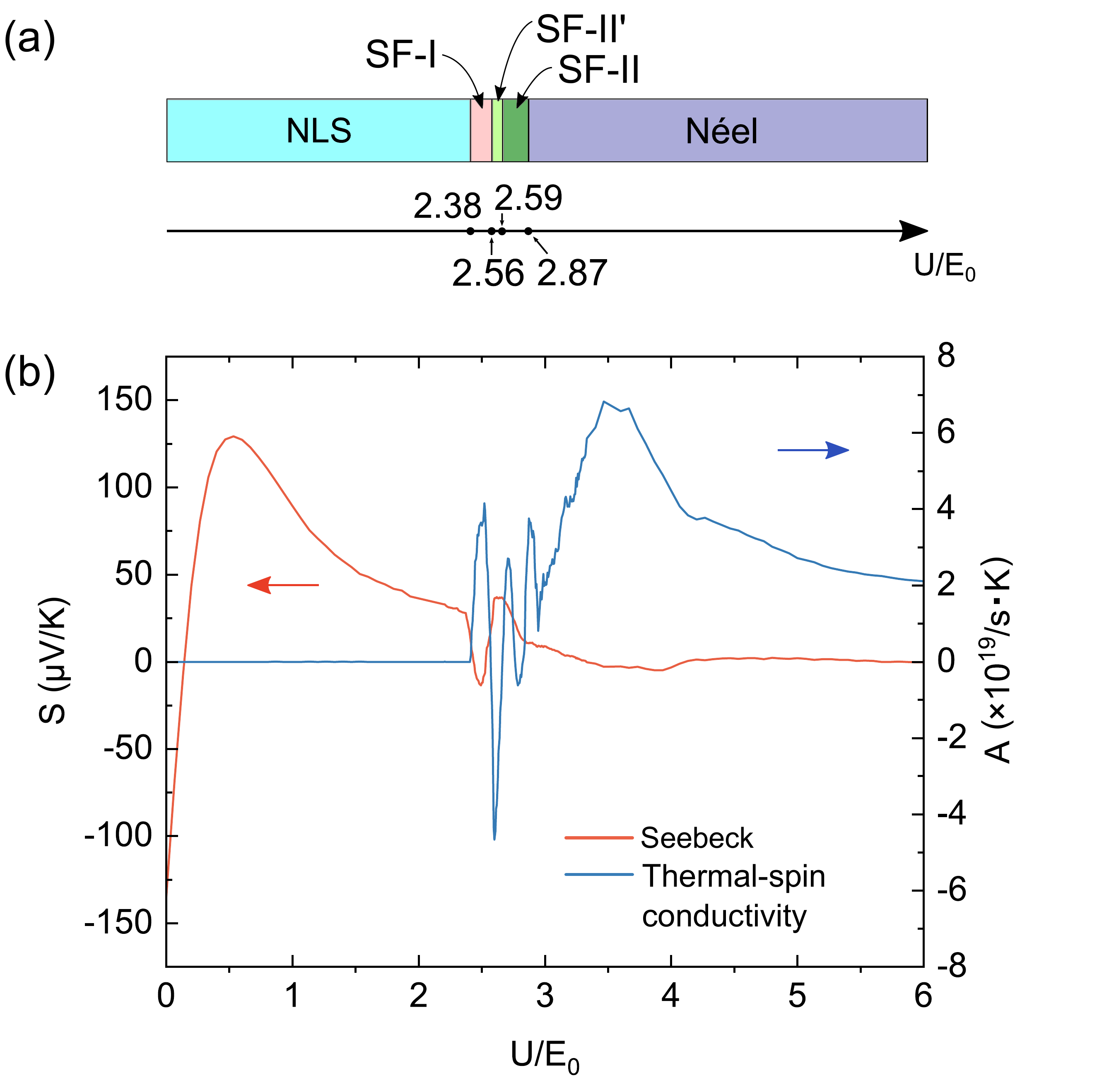}}
\caption{Effects of correlation on a TNLS film. (a) Phase diagram of a 10-layer TNLS film with type I truncation. SF denotes the surface ferromagnetic order, and NLS denotes the nodal line semimetal phase. The magnetic structures of each phase are shown in the Supplemental Materials. The horizontal axis is normalized by the one-body potential, $E_0$. (b) Seebeck and thermal-spin conductivities. The orange (blue) solid line corresponds to the left (right) axis and describes the results for the Seebeck coefficient (thermal-spin conductivity).}
\label{surfacemag}
\end{figure}

Starting from initial spin configurations with N$\acute{\mathrm{e}}$el and ferromagnetic order, we solve the self-consistent equations for the mean-field parameters $\langle n_{\ell\sigma}^\alpha\rangle$, where $\ell$ is a layer index. We determine the ground state by comparing the mean-field energy $\langle H_{\mathrm{mf}}\rangle$. We then calculate the magnetization for each layer to investigate the surface magnetism. For simplicity, we set $U_A=U_B=U$. As shown in Fig. \ref{surfacemag} (a), we find five phases as a function of $U$, including surface ferromagnetic (SF) orders. The detailed magnetic structures for each phase are provided in the Supplemental Materials. Note that the ground state is obtained from the initial state with N$\acute{\mathrm{e}}$el
for the full range of $U$. The phase boundaries are determined from the singular points of the second differentiation of the mean-field energy $\partial^2 \langle H_{\mathrm{mf}}\rangle/\partial U^2$. We note that a similar SF phase was found in a previous study for different TNLS models  \cite{liuCorrelationEffectsQuantum2017}.

Considering that magnetically polarized electrons support the spin current, we expect that the thermal-spin conductivity (defined below) will also by enhanced in SF phases. Note that the relation in Eq. \eqref{SB} holds even when Hubbard interaction is present \cite{kontaniGeneralFormulaThermoelectric2003}. The spin current ${\bm J}_s$ is defined as
\begin{equation}
  {\bm J}_s=\frac{1}{e}({\bm J}_e^\uparrow-{\bm J}_e^\downarrow)=\sum_{{\bm k},\sigma}\sigma\left(\frac{\partial \epsilon_{{\bm k}\sigma}}{\partial {\hbar\bm k}}\right)c_{{\bm k}\sigma}^\dagger c_{{\bm k}\sigma}.
\end{equation}
As in the case of the ordinary Seebeck effect, applying a condition where the electrical current is zero gives
\begin{equation}
  {\bm J}_s=\frac{1}{e}\frac{2(L_{11}^\uparrow L_{12}^\downarrow-L_{11}^\downarrow L_{12}^\uparrow)}{L_{11}^\uparrow+L_{11}^\downarrow}\frac{\nabla T}{T}\equiv A\nabla T,
\end{equation}
where we call $A$ the thermal-spin conductivity \cite{ogataTheorySpinSeebeck2017}.

The ordinary Seebeck coefficients and thermal-spin conductivity $A$ are shown in Fig. \ref{surfacemag} {(b)}. Note that the chemical potential is determined by the half-filled condition for each $U$, and the absolute value of $A$ is proportional to $\tau$, which is assumed to be $10^{-13}$ s in Fig. \ref{surfacemag}. There is a drastic sign change in $S$ when $U$ is small. This is because the one-body potential energy becomes layer-dependent owing to Coulomb interaction, so there is a shift in the energy of the state. In the SF phase, there is an oscillating finite thermal-spin conductivity and the positions of the sign changes correspond to phase transition points. Although the ordinary Seebeck coefficient shows a similar characteristic behavior, the thermal-spin conductivity is more sensitive to changes in magnetization. However, contrary to our expectations, the thermal-spin conductivity reaches its maximum in the N$\acute{\mathrm{e}}$el phase, rather than the SF phase. This is because the difference in the magnetizations of the A and B sublattices remained quite large in the region $3.0\lesssim U/E_0 \lesssim 4.0$. Note that, if the ferromagnetic phase is stable at large $U$, the thermal-spin conductivity would only be finite for the SF phases, as in the case of a one-dimensional quantum wire \cite{ogataTheorySpinSeebeck2017}. The construction of such a situation, and the material realization will be the subject of future works.

%%%%%%%%%%%%%%%%%%%%%%%%%%%%%%%%%%%%%%%%%%%%%%

\textit{Discussion and summary.}
In summary, we have shown that thin films of TNLSs are promising candidates for thermoelectric converters. The drumhead surface states result in a peculiar DOS structure, which means that, unlike typical semiconductors, the Seebeck coefficient is large and non-vanishing at the Fermi energy. Because the surface states are robust against disorder in the bulk, the phonon contribution to thermal conductivity can be reduced, which leads to a large $ZT$. In this work we found $ZT$ greater than 13 for type I 5-layer film. Remarkably, the chemical potential can be tuned easily by applying a gate voltage to the film, so the maximum $ZT$ could be found experimentally. Furthermore, we analyzed the correlation effect and determined the thermal-spin conductivity for finite $U$. However, methods of enhancing it are yet to be identified. Our results are restricted to a toy model and further investigation is needed to determine the extent to which the phonon contribution can be mitigated.

In real materials, Ag$_2$S is a good candidate for the realization of the present theoretical predictions, as the existence of an approximately flat drumhead surface states has been reported \cite{huangTopologicalNodallineSemimetal2017}. We expect that a similar situation occurs in this material, and a giant $ZT$ would be obtained for thin films.

\textit{Acknowledgments.}
This work was supported by Grants-in-Aid for Scientific Research from the JST-Mirai Program, Japan (grant Number JPMJMI19A1), and the Japan Society for the Promotion of Science (grant numbers JP20J10725, JP20K03802, JP18H01162, and JP18K03482). M. H. was supported by the Japan Society for the Promotion of Science through the Research Fellowship for Young Scientists and the Program for Leading Graduate Schools (MERIT).

\bibliography{nodal_paper_ref}

\end{document}

%%%%%%%%%% Merge with supplemental materials %%%%%%%%%%
\widetext
\clearpage
\begin{center}
  \textbf{\large
Supplemental Materials
  }
\end{center}
%%%%%%%%%% Merge with supplemental materials %%%%%%%%%%
%%%%%%%%%% Prefix a "S" to all equations, figures, tables and reset the counter %%%%%%%%%%
\setcounter{section}{0}
\setcounter{equation}{0}
\setcounter{figure}{0}
\setcounter{table}{0}
\setcounter{page}{1}
\makeatletter
\renewcommand{\thesection}{S-\Roman{section}}
\renewcommand{\theequation}{S\arabic{equation}}
\renewcommand{\thefigure}{S\arabic{figure}}
\renewcommand{\thetable}{S\arabic{table}}
\renewcommand{\bibnumfmt}[1]{[S#1]}
\renewcommand{\citenumfont}[1]{S#1}
%%%%%%%%%% Prefix a "S" to all equations, figures, tables and reset the counter %%%%%%%%%%

\section{I: Band structures of the type I\hspace{-1pt}I film and thermal transport properties}

\begin{figure}[h]
  \centering
\rotatebox{0}{\includegraphics[width=0.8\linewidth]{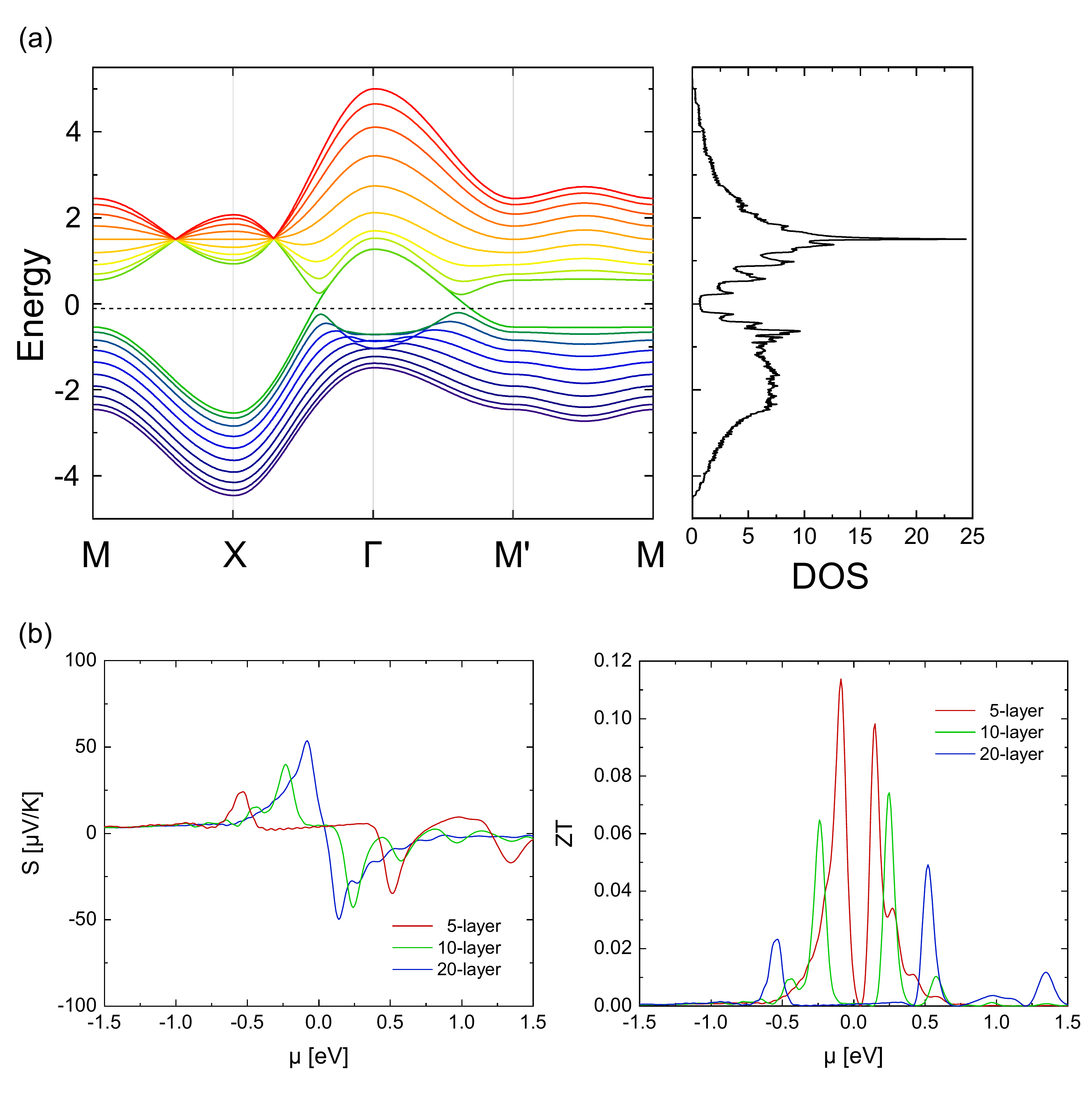}}
\caption{(a) Band structures of a 10-layer TNLS film with type I\hspace{-1pt}I truncation along the high symmetry points of the Brillouin zone. The coordinates of the high-symmetry points are given in the caption of Fig. 2. The dashed line represents the Fermi energy. The right panel shows the DOS. (b) Chemical potential dependence of the Seebeck coefficients $S$ and $ZT$ for the type I\hspace{-1 pt}I TNLS film. Red, green, and blue lines represent the results for 5-, 10-, and 20-layer films, respectively.}
\label{type2}
\end{figure}

This section describes the band structures for the type I\hspace{-1 pt}I truncation case in TNLS films, and examines their thermal transport properties.
The band structure for the type I\hspace{-1 pt}I case is shown in Fig. \ref{type2} (a). The projected bulk nodal line is a gapless point, as in the type I case. However, in contrast to the type I case, midgap surface states were not found. We can explain this difference from the perspective of the topological materials.

In our model, the band inversion at the $\Gamma$ point occurs between two bands with different parities, so the $\mathbb{Z}_2$ index \cite{ fukane2007topological, kim2015dirac} is calculated as $(1;111)$. This index represents a TNLS in a case where spin-orbit coupling is negligible, and there is a strong topological insulator for a strongly spin-orbit coupled case. In TNLSs, the $\mathbb{Z}_2$ index is directly related to the Zak phase. In this case, the Zak phases at the surface $\Gamma$ point and the surface X points differ by $\pi$. This difference in the Zak phase corresponds to the difference in the number of occupied bands at the $\Gamma$ and X points. A band localized at the surface \cite{vanderbilt1993electric}, actually crosses the Fermi energy near the gapless points. However, the Zak phase does not guarantee that the band will be a ``midgap" surface state. The detailed energy of the band depends on the surface parameters. In our thin film model, we used the same parameters as the bulk model, so the band is located at the upper edge of the bulk occupied band spectrum in the type I\hspace{-1 pt}I case.

However, in realistic materials, the surface parameters may change when the surface is made. Therefore, the surface state tends to appear as an isolated midgap state. For example, in our model, the type I\hspace{-1 pt}I film does not have the same number of A and B sites. In realistic materials, this corresponds to an imbalance between donors and acceptors. This can change the one-body potential of A site on the surface, and the band localized on the surface at the X point (and M point) can shift towards the Fermi energy. This is why drumhead surface states in TNLSs have been experimentally observed and reported in many previous studies, as explained in the main text.

The thermoelectric transport properties and figure of merit ($ZT$) for type I\hspace{-1 pt}I films with various thicknesses are shown in Fig \ref{type2} (b) and (c). The temperature was set to approximately room temperature, $k_BT=2.5\times10^{-2}$. In contrast to the type I case, the absolute value of $S$ and the layer dependence are small, which results in a banal value of $ZT$. However, in realistic materials, we can expect large values of $S$ and $ZT$, even for the type I\hspace{-1 pt}I case, because of the surface reconstruction described above.

\section{I\hspace{-1pt}I: Detailed derivation of $L_{11}$}
In this section, we present a detailed derivation of $L_{11}$. The electric conductivity $L_{11}$ under a uniform electric field is obtained from the equation
\begin{equation}
    L_{11}=\left.\frac{\Phi_{\mu\nu}(\omega)-\Phi_{\mu\nu}(0)}{i(\omega+i\delta)}\right|_{\omega\to0},
\end{equation}
where $\Phi_{\mu\nu}(\omega)=\left.\Phi_{\mu\nu}(i\omega_\lambda)\right|_{i\omega_\lambda\to\hbar\omega+i\delta}$. The quantities $\omega$ and $\omega_n$ are the frequency, and Matsubara frequency, respectively, and $i\omega_\lambda\to\hbar\omega+i\delta$ describes the analytic continuation.
If the vertex correction is negligible, then
\begin{equation}
  \begin{aligned}
    \Phi_{\mu\nu}(i\omega_\lambda)&=\frac{1}{V}\int_0^\beta d\tau\langle T_\tau\hat{j}_\mu(\tau)\hat{j}_\nu(0)\rangle e^{i\omega_\lambda\tau}\\
    &=-\frac{2e^2k_BT}{\hbar^2 V}\sum_{n,{\bm k}}\frac{\partial \epsilon_{\bm k}}{\partial k_\mu}{\mathcal{G}}({\bm k},i\epsilon_n-i\omega_\lambda)\frac{\partial \epsilon_{\bm k}}{\partial k_\nu}{\mathcal{G}}({\bm k},i\epsilon_n),
  \end{aligned}
\end{equation}
where the thermal Green's function within a constant relaxation time approximation is
\begin{equation}
{\mathcal{G}}({\bm k},i\epsilon_n)=\frac{1}{i\epsilon_n-\epsilon_{\bm k}+\mu+\frac{i\hbar}{2\tau}\text{sign}(\epsilon_n)}.
\end{equation}
Taking the Matsubara summation using a residue integral and analytical continuation $i\omega_\lambda\to\hbar\omega+i\delta$, we obtain
  \begin{equation}
    \begin{aligned}
      \Phi_{\mu\nu}(\omega)&=\frac{2e^2}{\hbar^2V}\sum_{\bm k}\frac{\partial \epsilon_{\bm k}}{\partial k_\mu}\frac{\partial \epsilon_{\bm k}}{\partial k_\nu}\times\\&\qquad\int_{-\infty}^\infty\frac{d\epsilon}{2\pi i}f(\epsilon)\left[G^R({\bm k},\epsilon)G^R({\bm k},\epsilon_+)-G^A({\bm k},\epsilon)G^R({\bm k},\epsilon_+)+G^A({\bm k},\epsilon_-)G^R({\bm k},\epsilon)-G^A({\bm k},\epsilon_-)G^A({\bm k},\epsilon)\right].
    \end{aligned}
  \end{equation}
Here, $G^A({\bm k},\epsilon)=G^{R*}({\bm k},\epsilon)$ and $\epsilon_\pm=\epsilon\pm\hbar\omega$. Therefore,
\begin{equation}
  \begin{aligned}
    L_{11}&=\frac{1}{i\omega}\frac{2e^2}{\hbar^2V}\sum_{\bm k}\frac{\partial \epsilon_{\bm k}}{\partial k_\mu}\frac{\partial \epsilon_{\bm k}}{\partial k_\nu}
    \int_{-\infty}^\infty\frac{d\epsilon}{2\pi i}f(\epsilon)(\hbar\omega)\left[(G^R-G^A)\frac{\partial}{\partial\epsilon}(G^R-G^A)\right]\\
    &=\frac{2e^2}{\pi\hbar V}\sum_{\bm k}\frac{\partial \epsilon_{\bm k}}{\partial k_\mu}\frac{\partial \epsilon_{\bm k}}{\partial k_\nu}\int_{-\infty}^\infty d\epsilon \left(-\frac{\partial f}{\partial \epsilon}\right)\left[{\mathrm{Im}}G^R({\bm k},\epsilon)\right]^2.
  \end{aligned}
\end{equation}
Inserting $G^R({\bm k},\epsilon)=1/(\epsilon-\epsilon_{\bm k}+i\hbar/2\tau)$ and executing a residue integral, finally gives
\begin{equation}
L_{11}=\frac{4\tau e^2}{\hbar^2V}\sum_{\bm k}\frac{\partial \epsilon_{\bm k}}{\partial k_\mu}\frac{\partial \epsilon_{\bm k}}{\partial k_\nu}(-f'(\epsilon_{\bm k})).
\end{equation}
%%%%%%%%%%%%%%%%%%%%%%%%%%%%%%%%%%%%%%
\section{I\hspace{-1pt}I\hspace{-1pt}I: Mean Field Approximation}
In the mean-field approximation, the Hubbard interaction is decoupled into
\begin{equation}
  H_U\sim \sum_{i,\alpha=A,B}U_\alpha(\langle n_{i\uparrow}^\alpha\rangle n_{i\downarrow}^\alpha+n_{i\uparrow}^\alpha\langle  n_{i\downarrow}^\alpha\rangle-\langle n_{i\uparrow}^\alpha\rangle \langle n_{i\downarrow}^\alpha\rangle).
\end{equation}
When we ignore particle number deviations within the same layer, the mean-field parameters are the layer, sublattice, and spin indices. Then, the values of the 4$N_\ell$ parameters are determined by a self-consistent calculation, where $N_\ell$ is the number of layers, and we set $N_\ell=10$.

Once the self-consistent equations are solved, we can define the magnetization for each layer and sublattice as
\begin{equation}
  M_\ell^\alpha=\langle n_{\ell\uparrow}^\alpha \rangle-\langle n_{\ell\downarrow}^\alpha \rangle,
\end{equation}
where $\ell$ is the index of the layer. Moreover, the full Hamiltonian can be divided into
\begin{equation}
  H=H_\uparrow\oplus H_\downarrow.
\end{equation}
Then, the thermoelectric transport coefficients $L_{ij}^\sigma$ are calculated using the energy spectrum of $H_\sigma$.
%%%%%%%%%%%%%%%%%%%%%%%%%%%%%%%%%%%%%%%
\section{IV: Magnetic structures for the five phases in a TNLS film with correlations}

\begin{figure}[H]
  \centering
\rotatebox{0}{\includegraphics[width=0.8\linewidth]{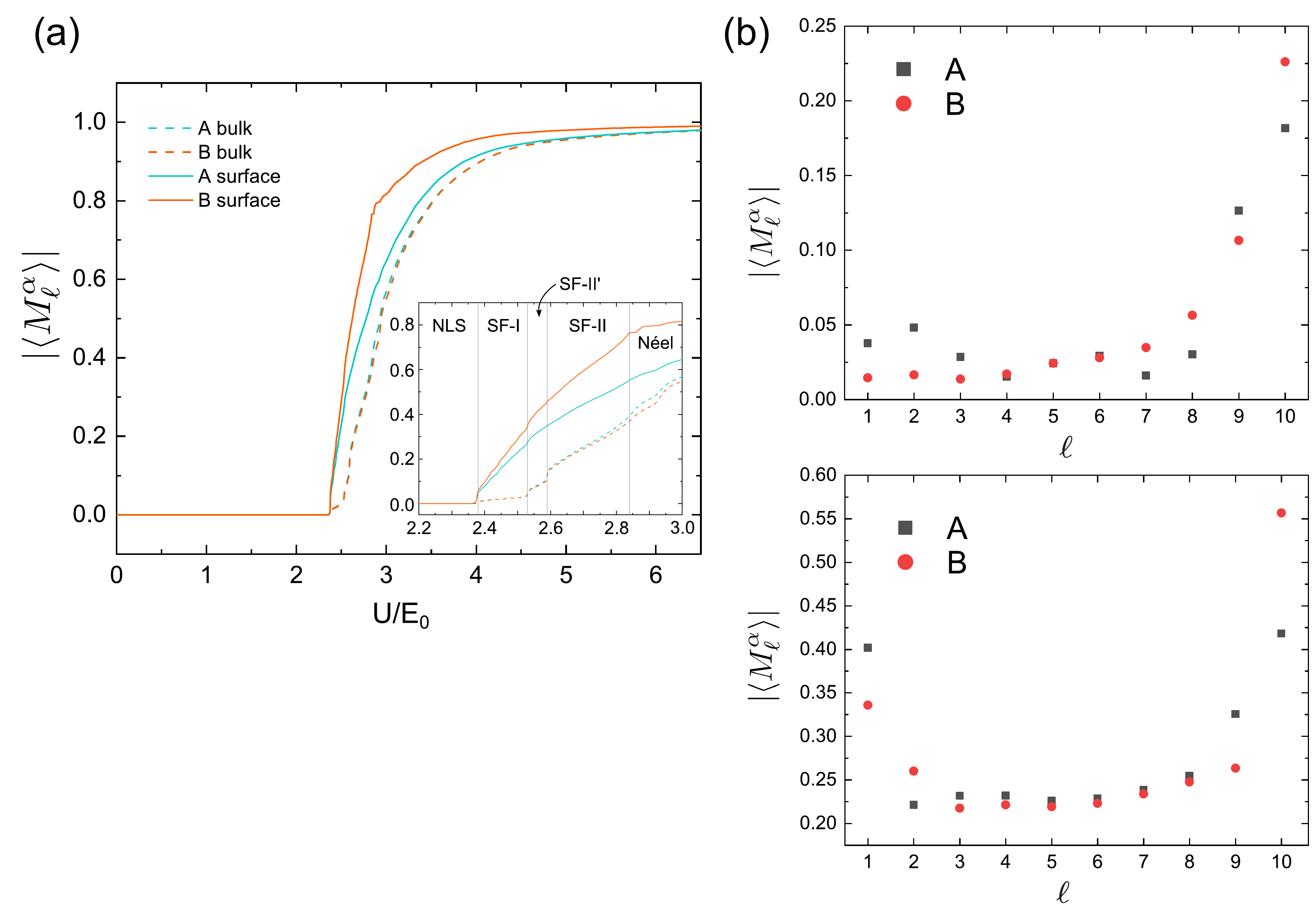}}
\caption{(a) Correlation $U$ dependence of the surface and bulk magnetization. The blue solid (dashed) line describes the result for the sublattice A layer at the surface (bulk), and the orange lines describe the corresponding results for the sublattice B layer. Here, we adopt $\ell=10$ as the surface layer and $\ell = 5$ as the bulk layer. The inset shows an extended view around the phase transition points, and the phase boundaries are also presented. (b) Magnetization of each layer. Top: SF-I phase ($U/E_0=2.47$). Bottom: SF-I\hspace{-1 pt}I phase ($U/E_0=2.67$). The black squares (red circles) show the magnetization of sublattice A (B).}
\label{surfacemag_detail}
\end{figure}

Fig. \ref{surfacemag_detail}(a) shows the correlation dependence of the surface and bulk magnetization.
When $U$ is small, the system remains in the nodal line semimetal state, and does not exhibit finite magnetization. Subsequently, it enters into a surface ferromagnetic phase, which we call the SF-I phase at $U/E_0=2.38$. In this phase, the magnetization of one side of the surface increases, and that of the other layers remains small (see the top panel of Fig. \ref{surfacemag_detail}(b)). Then, we encounter two phase transitions at $U/E_0=2.56,2.59$. Although the differences in the features of these two phases, which we name SF-I\hspace{-1 pt}I' and SF-I\hspace{-1 pt}I, respectively, are unclear, we find that the magnetization of both sides of the surface is more conspicuous than that of the bulk in the SF-I\hspace{-1 pt}I phase (see the bottom panel of Fig. \ref{surfacemag_detail}(b)). As $U$ increases, not only the surface magnetization but also the bulk magnetization grows, and the system enters into the N$\acute{\mathrm{e}}$el ordered phase at $U/E_0=2.87$. We note that even in this N$\acute{\mathrm{e}}$el ordered phase, the amplitude of magnetization for each layer is not uniform and a similar property of deviation to the SF phases remains, but the difference between the surface magnetization and the bulk magnetization is reduced.

\bibliography{nodal_appx_ref}